# Fermi Surface Study of Quasi-Two-Dimensional Organic Conductors by Magnetooptical Measurements

Yugo OSHIMA[*], Hitoshi OHTA[1], Keiichi KOYAMA[2], Mituhiro MOTOKAWA[2], Hiroshi M. YAMAMOTO[3], Reizo KATO[3], Masafumi TAMURA[3], Yutaka NISHIO[4] and Koji KAJITA[4]

*The Graduate School of Science and Technology, Kobe University, Rokkodai 1-1, Nada, Kobe 657-8501*

[1] *Molecular Photoscience Research Center, Kobe University, Rokkodai 1-1, Nada, Kobe 657-8501*

[2] *High Field Laboratory for Superconducting Materials, Institute for Materials Research, Tohoku University, Katahira, Sendai 980-8577*

[3] *RIKEN (The Institute of Physical and Chemical Research), Hirosawa, Wako 351-0198*

[4] *Faculty of Science, Toho University, Miyama, Funabashi 274-8501*

abstract

Magnetooptical measurements of several quasi-two-dimensional (q2D) organic conductors, which have simple Fermi surface structure, have been performed by using a cavity perturbation technique. Despite of the simple Fermi surface structure, magnetooptical resonance results show a dramatic difference for each sample. Cyclotron resonances (CR) were observed for θ-(BEDT-TTF)$_2$I$_3$ and (BEDT-TTF)$_3$Br(pBIB), while periodic orbit resonances (POR) were observed for (BEDT-TTF)$_2$Br(DIA) and (BEDT-TTF)$_3$Cl(DFBIB). The selection of the resonance seems to correspond with the skin depth for each sample. The effective mass of POR seems to have a mass enhancement due to the many-body effect, while effective mass of CR is independent of the strength of the electron-electron interaction. The scattering time deduced from each resonance's linewidth will be also presented.

KEYWORDS: magnetooptical measurements, periodic orbit resonance, cyclotron resonance, organic conductor, effective mass, linewidth, Fermi surface, cavity perturbation technique

---

[*] E-mail: yugo@phys.sci.kobe-u.ac.jp

§1. Introduction

Quasi-two-dimensional (q2D) organic conductors have attracted remarkable interest, because their physical properties change drastically with various anions or crystal structure. For this reason, the Fermi surface (FS) topologies of these salts have been studied by various high magnetic field technique.[1] Shubnikov-de Haas oscillation (SdH), de Haas-van Alphen oscillation (dHvA) and angle dependent magnetoresistance oscillation (ADMRO) are the most commonly used techniques for the FS studies. However, the magnetooptical measurement is another useful technique which is becoming a powerful tool for FS studies. Recent magnetooptical measurement studies have shown that the mechanisms responsible for the resonant absorption are quite different from the well-known conventional cyclotron resonance (CR). There are two interesting reports of harmonic resonances which are based on a simple FS structure. One is the harmonic cyclotron resonance (HCR), and the other is the periodic orbit resonance (POR). HCR predicts the presence of high-order CRs which are associated with the higher harmonics of oscillating real-space velocity of carriers in cyclotron orbits around the FS pockets, which does not predict even harmonics.[2] On the other hand, POR is arising from the periodic motion of the carriers in the anisotropic orbits of q2D FS which causes multiple resonance-like features in the interlayer conductivity where the second harmonic is dominant.[3]

We have previously reported magnetooptical measurements of α-(BEDT-TTF)$_2$KHg(SCN)$_4$ and observed four CRs in which each effective mass was smaller than those obtained by dHvA and SdH oscillation measurements.[4] This was pointed out earlier by Singleton *et al.* that this difference can be explained by the Kohn's theorem which suggests that the cyclotron resonance's mass is independent of the strength of the electron-electron interaction.[5,6] However, the FS structure of this salt is very complex due to the nesting of the FS at the density wave state. Therefore, we could not clarify which resonance corresponds to the FS's closed orbit. Thus, our strategy is to perform magnetooptical measurements of organic conductors, which have a simple FS, to determine the relationship between the cyclotron mass and Kohn's theorem, and see whether or not we can observe HCR or POR.

In this paper, we will report on the results of four BEDT-TTF organic conductors which have simple q2D FS. The First one is θ-(BEDT-TTF)$_2$I$_3$ which has a closed orbit at the zone boundary and a magnetic breakdown orbit (Fig. 1(a)). The second one is (BEDT-TTF)$_2$Br(DIA) which has only an anisotropic closed orbit (Fig. 1(b)). The third and the last one are (BEDT-TTF)$_3$Br(pBIB) and (BEDT-TTF)$_3$Cl(DFBIB) , respectively, which have also a closed orbit (Fig. 1(c)) but we notice that these are 3:1 compounds and the carrier density is different from the 2:1 salts. Thus, we can also examine how carrier density affects the effective mass. In the latter part in this paper, we will discuss briefly about the relation between the skindepth and the observed resonance, and about the scattering time deduced from the each resonance's linewidth. Finally, the effective mass for each resonance will be discussed.

§2. Experimental

We performed magnetooptical measurements using a cavity perturbation technique with a millimeter vector network analyzer (MVNA) and a 14 T superconducting magnet at IMR, Tohoku University. The experimental setup can be found in ref. 7 and we refer the reader to a series of articles on the cavity perturbation technique.[8-12] The typical sample size used for this study were from 0.25 x 0.5 x 0.1 to 1 x 1 x 0,1 mm$^3$ and the static magnetic field was applied perpendicular to the conducting plane. The cylindrical cavity was used for this study and the frequencies used are around 58 GHz and 72 GHz whose cavity mode corresponds to TE$_{011}$ and TE$_{012}$, respectively. If the sample is set in the middle of the cavity by using a polyethylene pillar, it has a different coupling mode by using different frequency, an *E*-field and *H*-field coupling when using around 58 GHz (TE$_{011}$) and 72 GHz (TE$_{012}$), respectively. However, if the sample is set on the end-plate of the cavity, the coupling mode around 58 GHz (TE$_{011}$) and 72 GHz (TE$_{012}$) will always be *H*-field coupling. We will hereafter call the former a pillar configuration, and the latter an end-plate configuration. In general, the *E*-field and *H*-field couplings are used for observing CR and ESR, respectively. However, we use both couplings for our magnetooptical measurements.

§3. Results and Discussion

*3.1 θ-(BEDT-TTF)$_2$I$_3$*

θ-(BEDT-TTF)$_2$I$_3$ is a q2D organic conductor without any phase transition at low temperatures.[13] The *ab*-plane is the conducting plane and there are two closed orbits where α and β orbits are attributed to the closed orbit around the zone boundary and the magnetic breakdown orbit, respectively (Fig. 1(a)). SdH measurements are already performed by Terashima *et al.* and the obtained effective masses for α and β orbits are 1.8m$_e$ and 3.5m$_e$, respectively.[13] Figure 2(a) shows typical cavity transmission spectra for pillar and end-plate configuration at around 57 GHz. The magnetic field is applied perpendicular to the *ab*-plane. We have observed one resonance at around 2.2 T for pillar configuration. However, no resonance was observed in the end-plate configuration. The mode coupling for around 58 GHz in the pillar configuration is the *E*-field coupling which means the observed resonance is CR. The reason why we did not observe even ESR signal in the end-plate configuration (i.e. *H*-field) is because the ESR linewidth is narrower than the instrumental resolution

(=5 mT).[14] Figure 2(b) shows spectra for the same pillar configuration but in higher fields up to 14 T. We notice that the x-axis of Fig. 2(b) is renormalized by the effective cyclotron mass in relative units of the free electron mass $m_e$ and the effective cyclotron mass can be obtained from the well-known equation,

$$m^* = \frac{eB}{2\pi\nu}, \quad (1)$$

where $B$ is the resonance field and $\nu$ is the observed frequency. We see here again the same CR at 1.0$m_e$ for 57.8 GHz but no other resonance was observed in higher field. However, we have observed an additional absorption line at 2.0$m_e$ for 103.5 GHz which suggests that this absorption needs a certain higher magnetic field to appear (i.e. magnetic breakdown orbit). Therefore, the observed CR at around 1.0$m_e$ and 2.0$m_e$ can be attributed to α and β orbits, respectively. If these resonances are really coming from α and β orbits which are q2D FS, the resonances' effective mass must have a 1/cosθ dependence, where θ is the angle between the magnetic field and the $c^*$-axis. Figure 3 shows each effective mass plotted by a function of 1/cosθ. Both CRs have a 1/cosθ dependence except the resonance of conducting electrons (CER) which is independent of the magnetic field direction. These results show that both CRs are coming from the q2D FS (i.e. α and β orbit). We show in Fig. 4 the frequency-field diagram of all observed CRs including our previous measurements by strip-line technique.[15,16] The solid lines are fitted by eq.(1) and the obtained effective masses for α and β orbits correspond to 1.0$m_e$ and 2.1$m_e$, respectively.

Next, let us direct our attention to the gap affording the magnetic breakdown effect. The breakdown field $B_{MB}$ is approximately related to the energy gap near the junction $E_g$ by the equation,

$$B_{MB} = \frac{2\pi m^* E_g^2}{he\varepsilon_F}, \quad (2)$$

where $\varepsilon_F$ is the Fermi energy.[17] The breakdown field seem to be around 5 T from our results in Fig. 4. If we assume the Fermi energy $\varepsilon_F \sim 0.1$ eV and the effective mass $m^* \sim 2m_e$, the energy gap corresponds to $E_g \sim 5.4$ meV which is consistent with the topology of the FS deduced from the tight-binding band model and the dHvA results.[18,19] However, the breakdown field in magnetooptical measurements is much lower than that in SdH measurements which is around 10 T.[13] The intensity of CR is depending on the scattering time and the carrier density. And it seems that such a high field of

10 T is not needed to observe the β orbit in CR measurements due to the high carrier density and longer scattering time for β orbit as we will show in §4.

*3.2 (BEDT-TTF)$_2$Br(DIA)*

(BEDT-TTF)$_2$Br(DIA) is a novel q2D conductor with a very simple FS (Fig. 1(b)). One of the interesting features of the structure is that the supramolecular ...Br...DIA... chains are formed and the donor molecules fit into the channels formed by one dimensional (1D) chains along the *a-c* direction.[20] SdH and ADMRO measurements have already been performed and the results indicate the presence of a q2D FS with an elliptic cross-sectional area and the effective mass is 4.3m$_e$.[21]

Figure 5 shows typical spectra of (BEDT-TTF)$_2$Br(DIA) at 0.5 K in the pillar configuration. Here again, the x-axis is plotted by the effective mass in relative units of m$_e$. At each frequency, we can observe several harmonic absorption lines which become larger as the field increases. For harmonic resonances, the resonance condition will be $\omega=n\omega_c$, where $\omega_c$ is the cyclotron frequency and $\omega$ is the microwave frequency, which cause *n*th-order higher harmonics. Therefore, these harmonic absorption lines appear periodically as a function of the inverse field, and the effective mass can be obtained by the periodicity of the resonances from the condition mentioned above.[22] The estimated effective mass for example at 72 GHz would be around 4.4m$_e$. Then by reevaluating the resonance condition, the order of higher harmonics correspond to the vertical broken lines described in Fig. 5.

Figure 6 show the typical spectra at 0.5 K in the end-plate configuration. In general, the *H*-field coupling is the ideal configuration for observing ESR signals. Thus ESR signals are clearly observed in this configuration. The strongest absorption at 2.7m$_e$ corresponds to the second harmonic and the n=1 resonance is observed at around 4.9m$_e$. Both results in Figs. 5 and 6 suggest that the second harmonic is dominant which is the characteristic feature of POR. POR only appears when the FS shape is an anisotropic one which is in a good agreement with the ADMRO result.[21] However, the end-plate configuration shows the characteristic behavior of POR clearer than the pillar configuration, and this can be explain as follows. When the sample is set on the end-plate, the induced current excited from the oscillatory *H*-field rounds the sample surfaces.[23] However, the in-plane and the interlayer conductivity differ considerably for q2D organic conductor. Thus, the interlayer skin depth will be larger than the in-plane skin depth which means that the interlayer ac magnetoconductivity is mainly probed for this kind of measurements. This may be why the POR is clearer in the end-plate configuration. On the other hand, higher order of harmonics (i.e. 5th, 6th,

7th...) was not observed in the end-plate configuration. The in-plane conductivity in pillar configuration mode might play an important role for observing higher order of harmonics.

*3.3 (BEDT-TTF)$_3$Br(pBIB), (BEDT-TTF)$_3$Cl(DFBIB)*

(BEDT-TTF)$_3$Br(pBIB) and (BEDT-TTF)$_3$Cl(DFBIB) are in the same family of (BEDT-TTF)$_2$Br(DIA) which contains supramolecular assemblies. These supramolecular assemblies formed an 1D chains and the donor molecules (i.e. BEDT-TTF) fit into the channels formed by 1D chains which may suggest the possibility of "fractional band-filling control" by changing 1D supramolecular assemblies with a different period. (BEDT-TTF)$_3$Br(pBIB) and (BEDT-TTF)$_3$Cl(DFBIB) are one of these interesting salts which has the 3:1 donor/anion ratio achieved by a longer 1D supramolecular chain period.[24] These salts are the first metallic BEDT-TTF salts where formal charge of donor is +1/3 and the tight-binding calculation shows that the FS is composed of an anisotropic one (Fig. 1(c)).[24] The effective masses of Br(pBIB) and Cl(DFBIB) salts are obtained from SdH measurements and correspond to 2.0m$_e$ and 2.1m$_e$, respectively.[25] However, it is very interesting to perform the magnetooptical measurements in these salts and to investigate how carrier density affects the effective mass compared to 2:1 compounds.

Figure 7 represents the typical spectra of (BEDT-TTF)$_3$Br(pBIB) at 1.4K for 57.2 GHz in the pillar configuration. Only one absorption line is observed and can be assigned to the CR of the elliptic q2D FS from the *E*-field coupling of the cavity mode. However, no signals of POR were observed in the other frequency or sample configuration. The effective mass of (BEDT-TTF)$_3$Br(pBIB) is estimated as 1.0m$_e$.

Next, we show the typical spectra of (BEDT-TTF)$_3$Cl(DFBIB) at 0.5 and 4.2 K in the pillar configuration in Fig. 8. The frequency is around 72 GHz and the two spectra at each temperature show upward and downward scans. In the spectra of 4.2 K, we see a gradient due to its magnetoresistivity. However, we see some additional small absorption lines at 0.5 K. To show clearly the additional absorption lines and get rid of the gradient, we have divided several 0.5 K spectra by 4.2 K spectra which are shown in the inset of Fig. 8. The harmonic absorption lines are clearly seen and the absorption lines are reproduced in each spectrum. The second harmonics are the strongest one which suggests that observed absorption lines are POR.[3] Normally, the fundamental POR (i.e. *n*=1) cannot be observed when the magnetic field is applied perpendicular to the plane.[3] However, as in the (BEDT-TTF)$_2$Br(DIA) case, we have observed *n*=1 resonance whose effective mass corresponds to 2.9m$_e$. The reason why the fundamental POR is observed may be due to the sample misalignment or tilting of FS. The observation of POR is in a good agreement with the band calculation and ADMRO results which suggest anisotropic q2D FS.[24,25]

§4. CR, POR and resonance linewidths

Table I shows all the presented results with effective masses compared with SdH's effective mass results and the sample's skin depth for interlayer current at 60 GHz. Our presented results show that CRs were observed for θ-(BEDT-TTF)$_2$I$_3$ and (BEDT-TTF)$_3$Br(pBIB) and PORs were observed for (BEDT-TTF)$_2$Br(DIA) and (BEDT-TTF)$_3$Cl(DFBIB). No sample shows the observation of both CR and POR at the same time. Generally speaking, CR is the resonance related with the real part of the in-plane conductivity. However, POR is related with the interlayer conductivity which means the dissipation of POR is proportional to the skin depth for interlayer current. As shown in Table I, the skin depth for interlayer current differs considerably for each sample. The sample, in which CRs were observed, has a skin depth of 20-40 μm while the sample, in which POR were observed, has a skindepth of hundreds of μm. Thus, the skindepth for interlayer current may play an important role for the POR observation. A sample, which has a skindepth between 40-100 μm (e.g. α-(BEDT-TTF)$_2$KHg(SCN)$_4$), might have the opportunity to observe both CR and POR.

Finally, we would like to end this section by discussing the resonance linewidths. The resonance linewidth is governed by the scattering time $\tau$, nevertheless if it is CR or POR. Typical scattering time of θ-(BEDT-TTF)$_2$I$_3$ deduced from the linewidth data indicates the values of $\omega\tau$ between 10-14, i.e. $\tau \sim 1.5 \times 10^{-11}$-$3.6 \times 10^{-11}$ s. This results is an order of magnitude larger than the value obtained by dHvA measurements (i.e. $1.5 \times 10^{-12}$ and $0.6 \times 10^{-12}$ s for α and β orbit, respectively).[19] The scattering time may contain many different contributions from impurities, electron-phonon scattering, electron-electron scattering, etc. and the scattering time of CR/POR is not necessarily to be the same with the value of dHvA or SdH.[3,17] This scattering time difference may be the reason why CR/POR is observed at higher temperature and lower magnetic field than the dHvA and SdH measurements. The scattering time shows a considerable sample dependence and frequency dependence (i.e. magnetic field dependence). However, the scattering time $\tau$ for θ-(BEDT-TTF)$_2$I$_3$ at 103.5 GHz, 0.5K are $2.0 \times 10^{-11}$ and $1.7 \times 10^{-11}$ s for α and β orbit, respectively. (BEDT-TTF)$_3$Br(pBIB), in which CR was observed, had a similar results of $2.7 \times 10^{-11}$ s at 57.2 GHz, 1.4K. On the other hand, POR linewidths have the value of $\omega\tau$ between 2-3 which have a shorter scattering time than that of CR; $\tau \sim 4.6$-$5.5 \times 10^{-12}$ s and $5.6 \times 10^{-12}$ s for (BEDT-TTF)$_2$Br(DIA) and (BEDT-

TTF)$_3$Cl(DFBIB), respectively. We note that the linewidths for POR is evaluated from the second harmonics because the *n*=1 resonance linewidth is not so clear. However, the scattering time for POR is still longer than that from SdH result (i.e. $\tau$~1.7x10$^{-12}$ s).[25] These differences among CR, POR and SdH scattering times are very interesting, and these sample dependence of skin depth and scattering time might be the key for observing CR and POR and even SdH oscillation in magnetooptical measurements.

§5. CR and POR effective masses

Finally, we will discuss about the effective masses in this section. The effective mass for CRs are around 1.0m$_e$ except the magnetic breakdown orbit (i.e. β orbit) of θ-(BEDT-TTF)$_2$I$_3$ which is 2.1m$_e$. However, the CR effective mass is about a half smaller than those of the SdH measurements which suggests the Kohn's theorem is applicable to these salts. Kohn's theorem shows that for an isotropic three dimensional metal the effective mass is independent of electron-electron interaction. However, from our results, it seems that it is applicable also for a two-dimensional system, and from the comparison with SdH results, there exists not negligible electron-electron interaction in the system. On the other hand, POR's effective masses are similar with or slightly larger than those of SdH measurements. The Kohn's theorem is based on a three dimensional metal with the electric field perpendicular to the magnetic field which can be converted to the in-plane conductivity perpendicular to the magnetic field for a two dimensional system (i.e. CR). However, POR is related to the interlayer conductivity 'parallel' to the magnetic field as mentioned above. In that case, the Kohn's argument may break down and it is not clear whether Kohn's theorem is applicable to the interlayer conductivity. Furthermore, Kanki and Yamada have calculated the effects of electron-electron interaction on cyclotron resonance frequency on the basis of the Fermi liquid theory.[26] They suggest that the cyclotron masses strongly depend on the characteristics of the material (e.g. band-filling, symmetry of the FS) and in some extreme situations such as near half-filling on square lattice, cyclotron effective masses can be comparable with the effective mass obtained by SdH or dHvA measurements. However, they are considering only the CR's case (i.e. in-plane conductivity). Therefore, without further work, it is not clear if it is applicable to the interlayer conductivity. In particular, it is quite possible that the effective mass of POR is enhanced due to the many-body effect in contrary to CR's case. On the other hand, the POR effective mass for (BEDT-TTF)$_2$Br(DIA), 4.7m$_e$, is much larger than the mass for (BEDT-TTF)$_3$Cl(DFBIB) which is 2.9m$_e$. This difference may due to the carrier density difference of 2:1 and 3:1 compounds. However we need a more precise discussion and it remains as a subject for the future.

§6. Conclusion

Magnetooptical measurements of several quasi-two-dimensional (q2D) organic conductors, which have simple Fermi surface structure, have been performed by using a cavity perturbation technique. CRs were observed for θ-(BEDT-TTF)$_2$I$_3$ and (BEDT-TTF)$_3$Br(pBIB) and the effective mass is about a half smaller than those obtained from SdH measurements, which suggest not negligible electron-electron interaction exists in the system. On the other hand, POR were observed for (BEDT-TTF)$_2$Br(DIA) and (BEDT-TTF)$_3$Cl(DFBIB). Obtained effective mass is consistent with SdH measurements, which suggests a mass enhancement for POR's effective mass due to the many-body effect. The selection of CR or POR seems to correspond with the skin depth for interlayer current of each sample.

Acknowledgements

This work was supported by Grant-in-Aid for Scientific Research (B) (No. 10440109), and Grant-in-Aid for Scientific Research on Priority Areas (A) (No. 11136231, 12023232 Metal-assembled Complexes) from the Ministry of Education, Culture, Sports, Science and Technology of Japan. This work has been performed at High Field Laboratory for Superconducting Materials, Institute for Materials Research, Tohoku University.

Figure Captions

Fig. 1.
Fermi surface of (a) θ-(BEDT-TTF)$_2$I$_3$, (b) (BEDT-TTF)$_2$Br(DIA), (c) (BEDT-TTF)$_3$Br(pBIB) and (BEDT-TTF)$_3$Cl(DFBIB).

Fig. 2.
(a) Typical cavity transmission spectra of θ-(BEDT-TTF)$_2$I$_3$ for pillar and end-plate configuration at around 57 GHz and 1.8 K.
(b) Typical spectra of θ-(BEDT-TTF)$_2$I$_3$ for pillar configuration with different frequencies at 1.9 K. The x-axis is renormalized by the effective cyclotron mass in relative units of free electron mass.

Fig. 3.
Angular dependence of α and β orbit's cyclotron effective mass for θ-(BEDT-TTF)$_2$I$_3$ at 1.9 K.

Fig. 4.
The frequency-field diagram of cyclotron resonances for θ-(BEDT-TTF)$_2$I$_3$ when the magnetic field is applied perpendicular to the *ab*-plane.

Fig. 5.
Typical cavity transmission spectra of (BEDT-TTF)$_2$Br(DIA) observed at 0.5 K in the pillar configuration. The vertical broken lines show the order of harmonics. The magnetic field is applied perpendicular to the *a\*c\**-plane.

Fig. 6.
Typical cavity transmission spectra of (BEDT-TTF)$_2$Br(DIA) observed at 0.5 K in the end-plate configuration. The triangles show ESR and the vertical broken lines show the order of POR's harmonics. The magnetic field is applied perpendicular to the *a\*c\**-plane.

Fig. 7.
Typical cavity transmission spectra of (BEDT-TTF)$_3$Br(pBIB) observed at 1.4 K in the pillar configuration. The magnetic field is applied perpendicular to the *a\*c\**-plane.

Fig. 8.
Typical cavity transmission spectra of (BEDT-TTF)$_3$Cl(DFBIB) observed at 0.5 and 4.2 K. The inset shows the spectra of 0.5 K divided by 4.2 K spectra. The magnetic field is applied perpendicular to the *a\*c\**-plane.

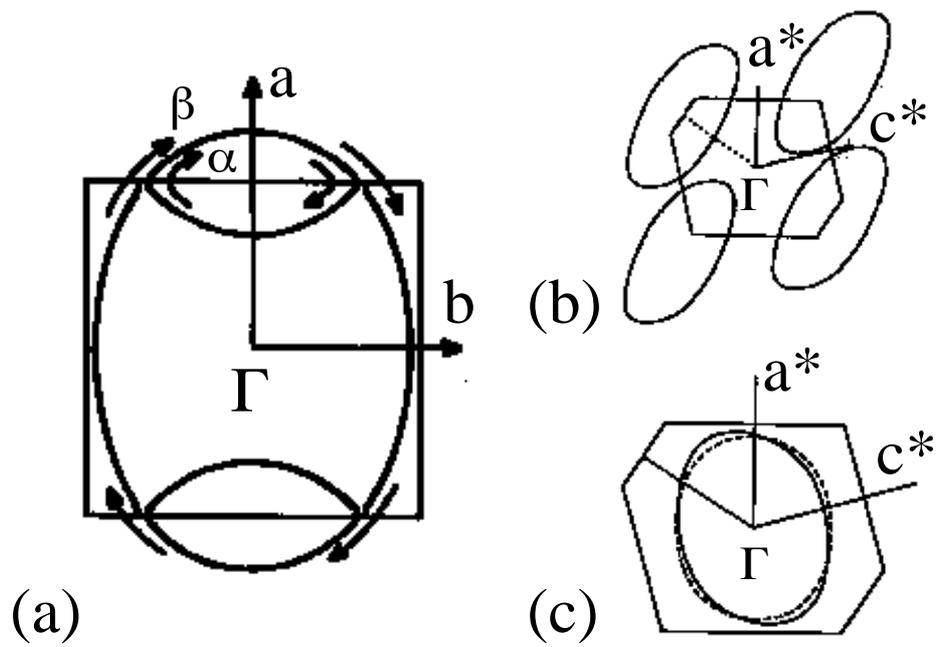

Fig. 1 (Size7~8cm)
Full Paper No.60163   Y. Oshima et al.

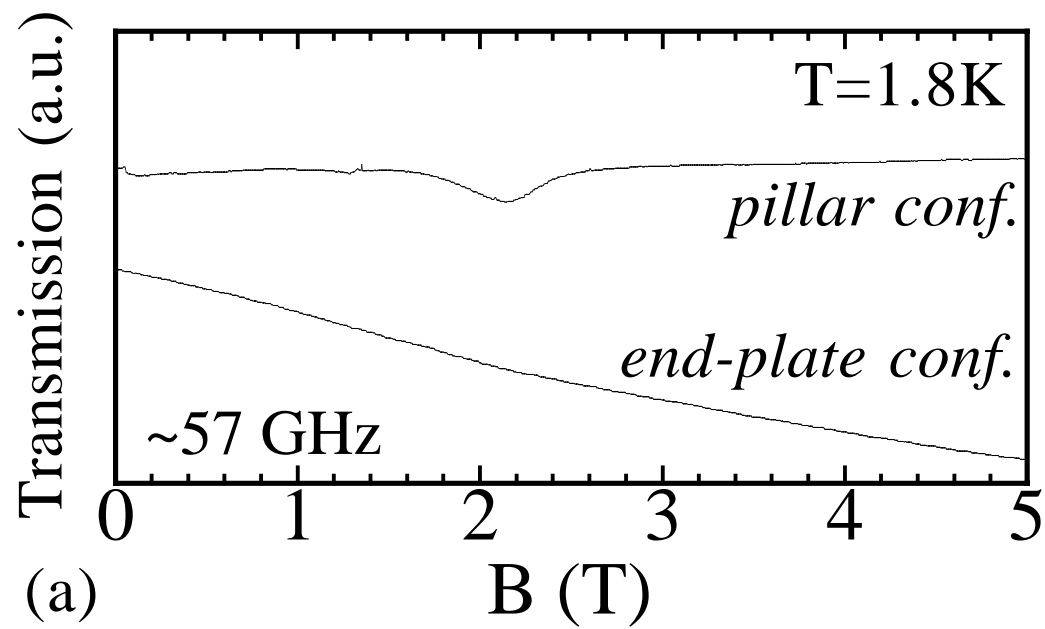

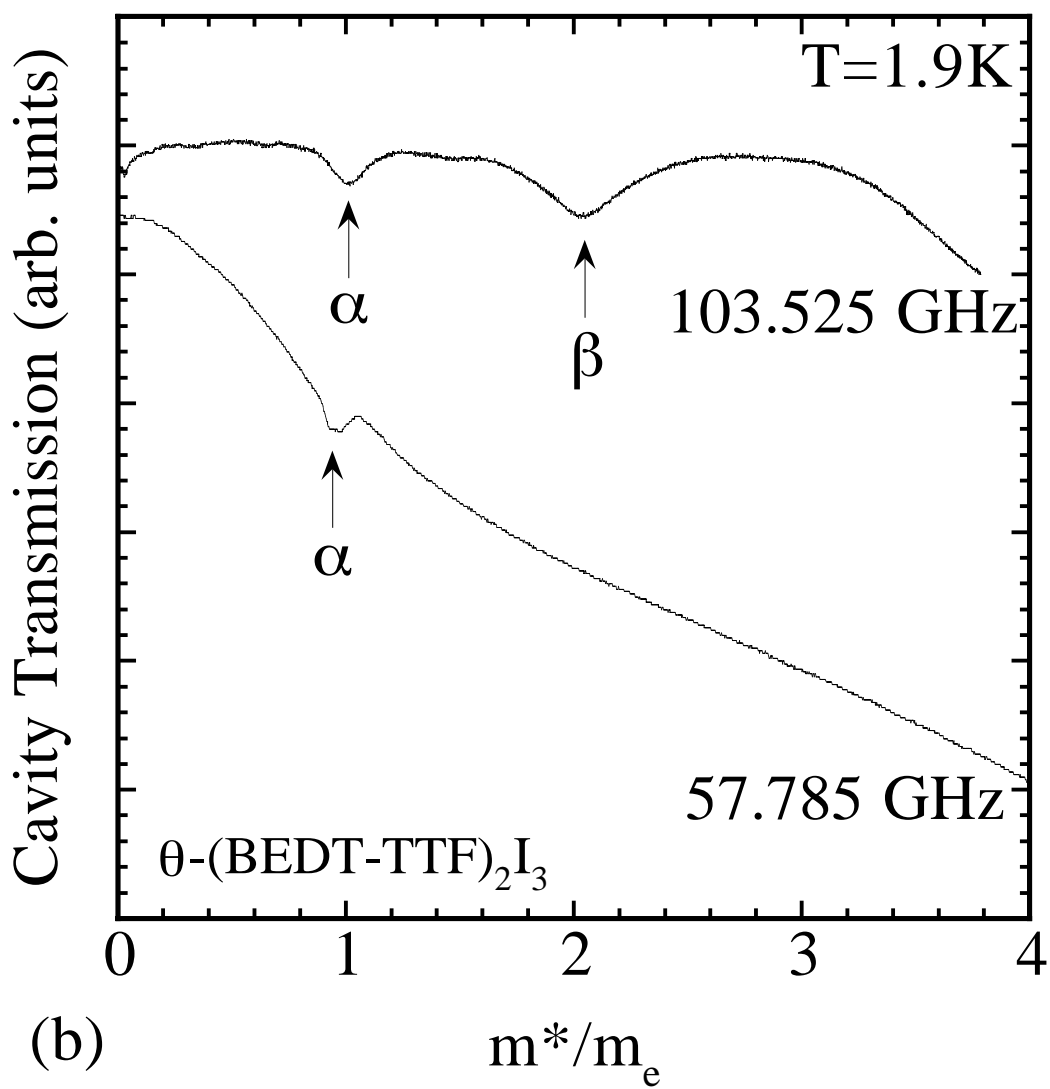

Fig.2 (size 7~8cm)　　Full Paper No.60163　Y. Oshima et al.

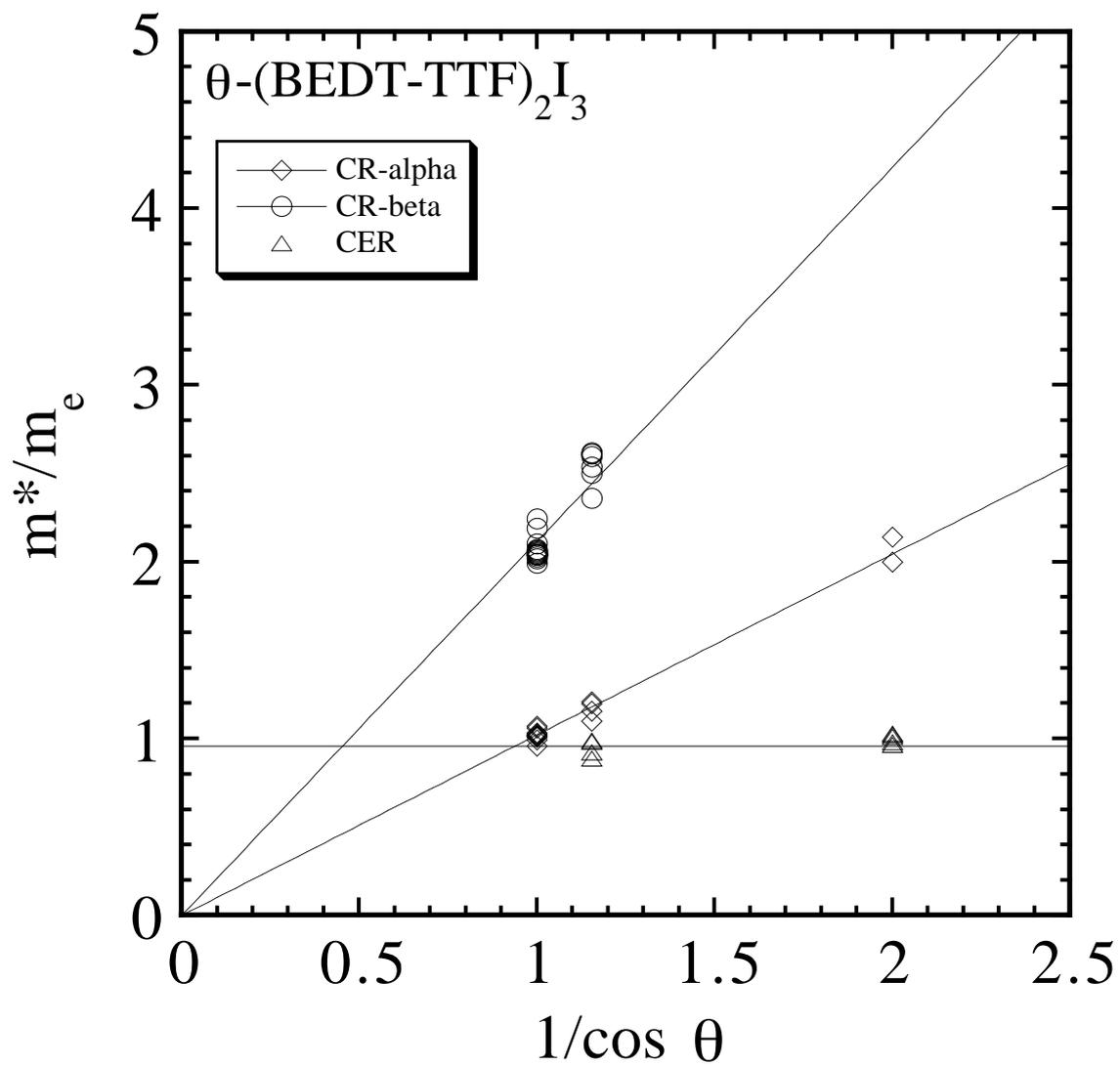

Fig. 3 (Size7~8cm)
Full Paper No.60163  Y. Oshima et al.

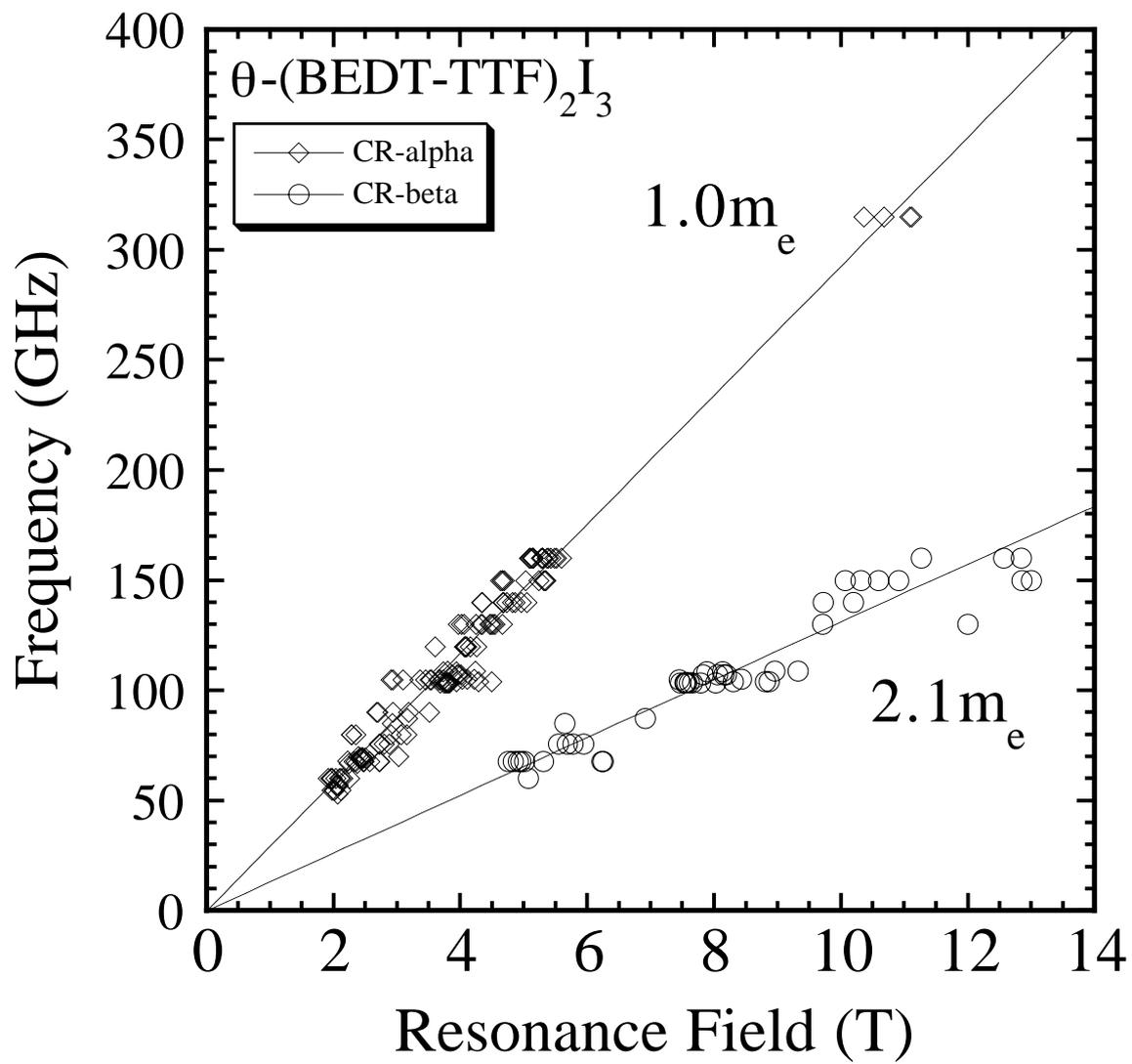

Fig. 4 (Size7~8cm)
Full Paper No.60163  Y. Oshima et al.

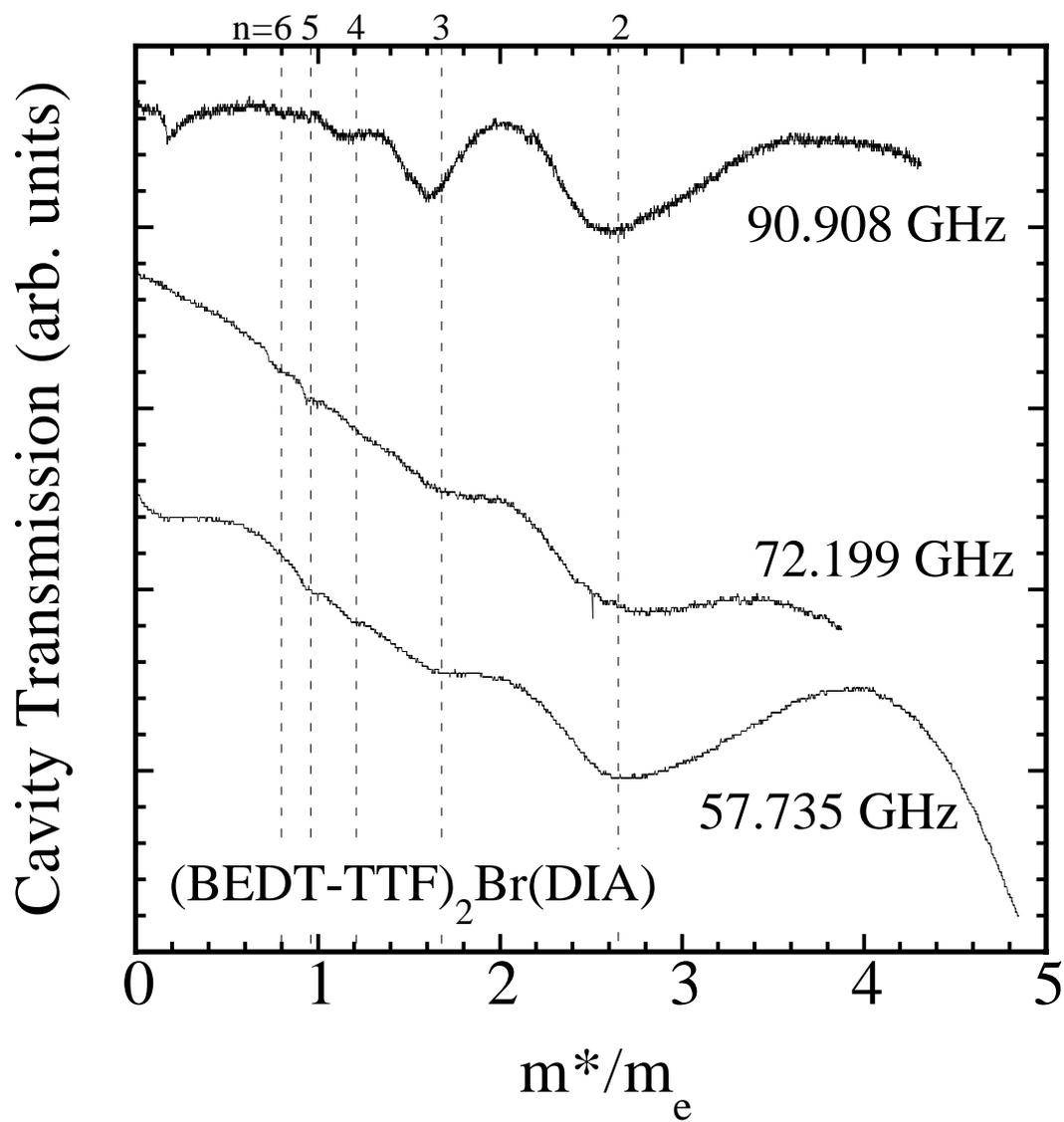

Fig. 5 (Size7~8cm)
Full Paper No.60163  Y. Oshima et al.

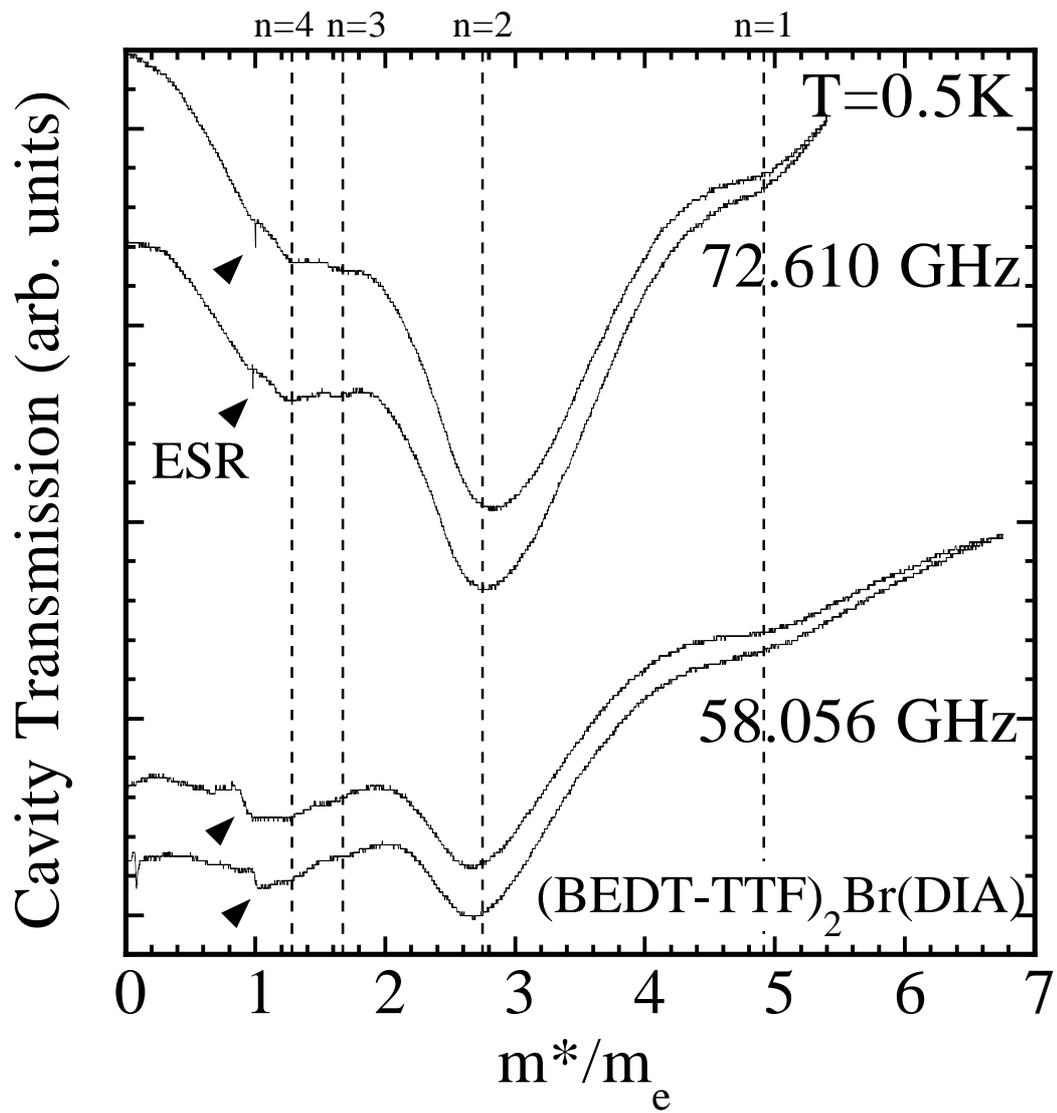

Fig. 6 (Size7~8cm)
Full Paper No.60163  Y. Oshima et al.

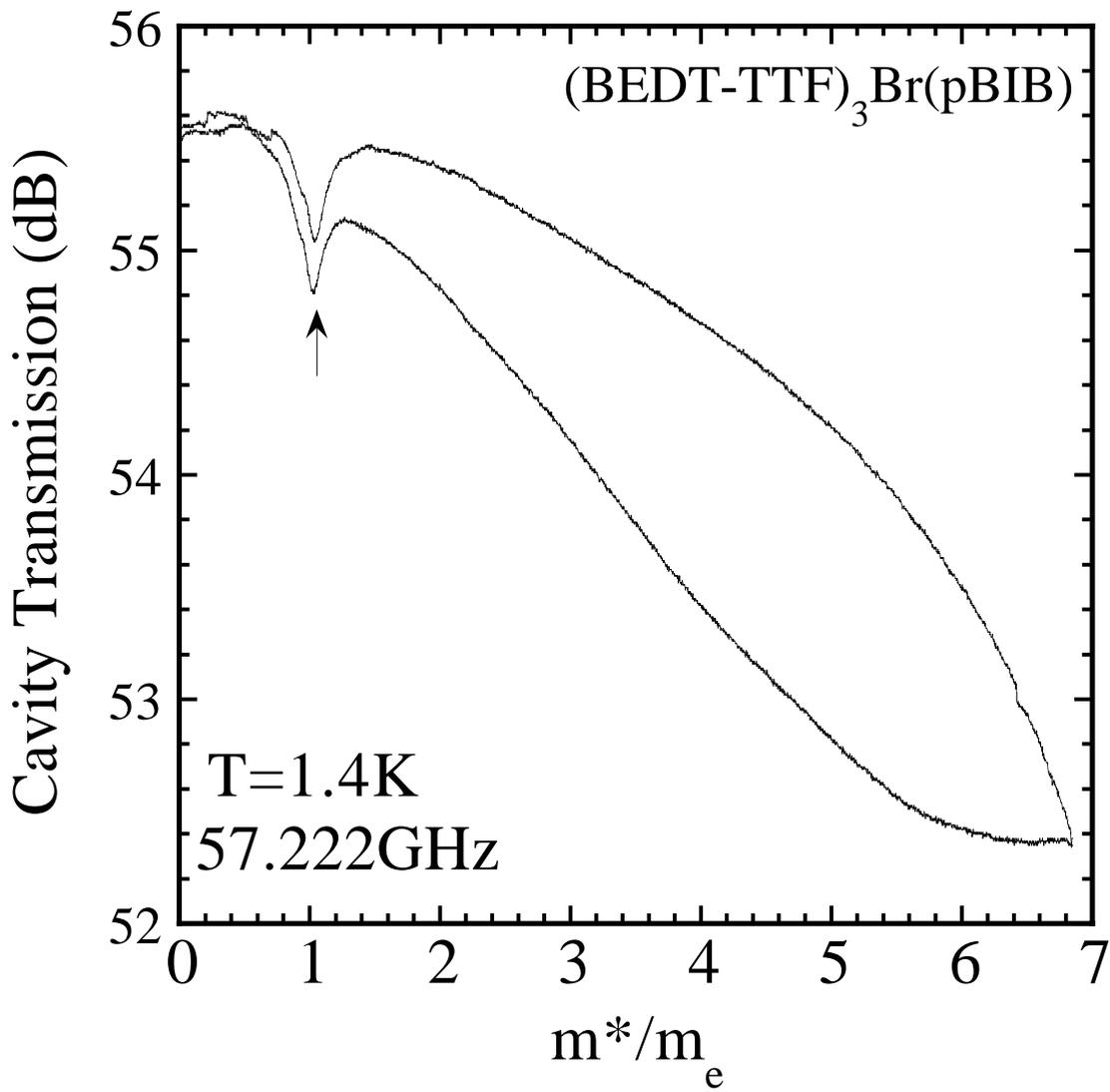

Fig. 7 (Size7~8cm)
Full Paper No.60163  Y. Oshima et al.

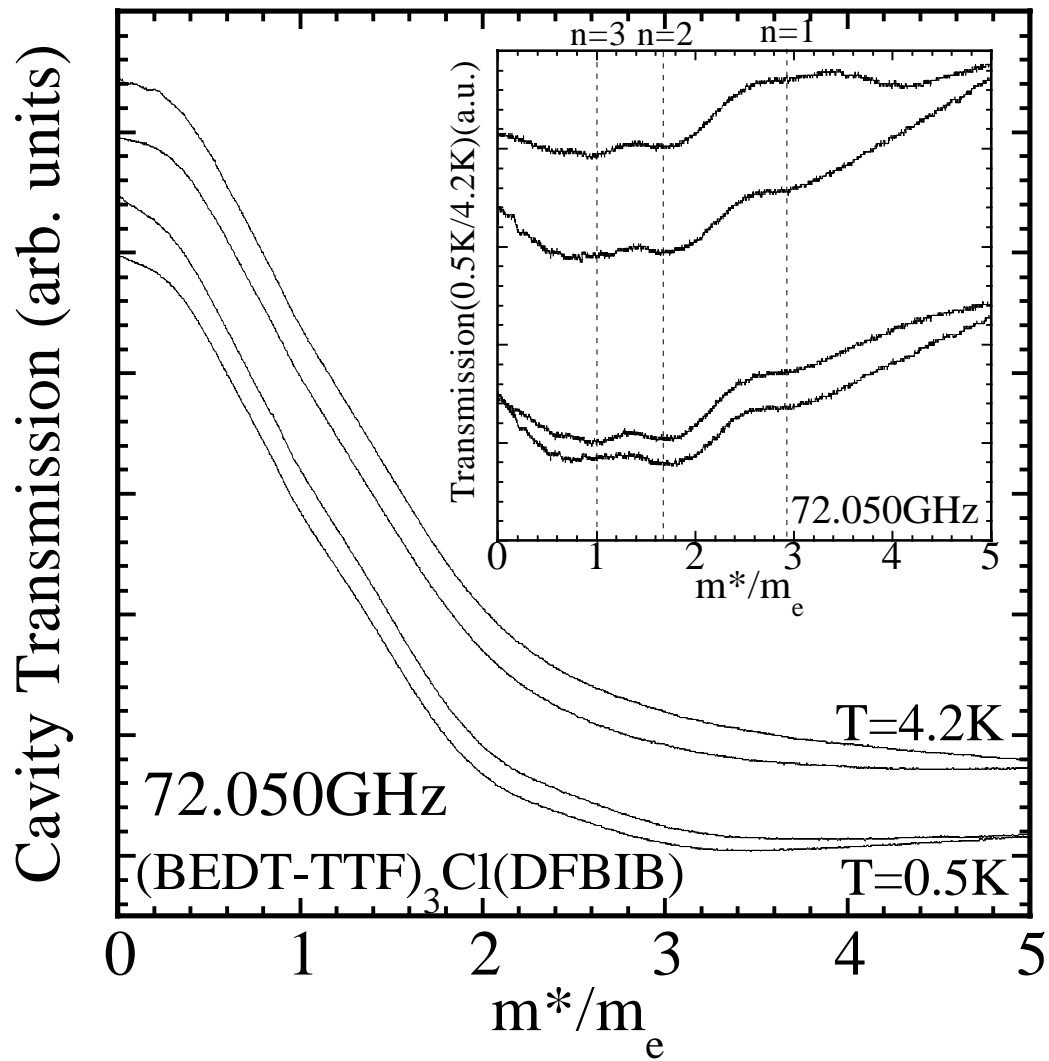

Fig. 8 (Size7~8cm)
Full Paper No.60163   Y. Oshima et al.

Table I. Our results with effective masses compared with SdH's results and sample's skin depth for interlayer current at 60 GHz, 4.2 K.

|  | CR | POR | SdH | Skindepth(60GHz) |
|---|---|---|---|---|
| $\theta$-ET$_2$I$_3$ | $\alpha$:1.0m$_e$ $\beta$:2.1m$_e$ | No POR | $\alpha$:1.8m$_e$ $\beta$:3.5m$_e$ [13] | 40μm |
| ET$_2$Br(DIA) | No CR | 4.7m$_e$ | 4.3m$_e$ [21] | 200μm |
| ET$_3$Br(pBIB) | 1.0m$_e$ | No POR | 2.0m$_e$ [25] | 20μm |
| ET$_3$Cl(DFBIB) | No CR | 2.9m$_e$ | 2.1m$_e$ [25] | 145μm |

ET=BEDT-TTF

Full Paper No.60163   Y. Oshima et al.